\documentclass[twocolumn,showpacs,preprintnumbers,amsmath,amssymb,floatfix]{revtex4}

\usepackage{graphicx}
\usepackage{dcolumn}
\usepackage{bm}

\begin{document}


\title{A consistent formalism for the Thomas-Ehrman Level Displacement}

\author{J.J.~He}
\email{hjianjun@ph.ed.ac.uk}
\author{A.St.J.~Murphy}

\affiliation{School of Physics, University of Edinburgh, Mayfield Road, Edinburgh, EH9 3JZ, United Kingdom}

\date{\today}  

\begin{abstract}
Usage of the Thomas-Ehrman Level Displacement formalism has been examined. 
Mistakes and inconsistencies are found in several papers, being repeated
in subsequent works. Here, we present a complete formalism with a consistent set of definitions. Full 
algorithms are made available, both as a {\tt FORTRAN} source file 
and as a user-friendly Visual Basic executable tool, available for download 
on the World Wide Web.

\end{abstract}


\maketitle

\section{Introduction} 

The Thomas-Ehrman Level Displacement formalism (TELD)~\cite{bib:ehr51,bib:tho52} is an established technique for calculating the level displacement between mirror pairs. It 
is found to be particularly useful in situations where a reaction proceeds via a proton resonant state in a proton-rich nucleus.
Largely, this usefulness derives from the fact that such states are above the particle decay threshold, usually resulting in proton partial
widths too narrow to be measured experimentally. Thus, by appealling to the charge symmetry of the nuclear force, one may make use of
relatively abundant spectroscopic data of analogue states in the mirror nucleus to determine the properties of the astrophysically interesting states. 
Examples in the literature are the $^{21}$Ne-$^{21}$Na~\cite{bib:mar57}, $^{20}$F-$^{20}$Na~\cite{bib:lan86}, 
$^{18}$O-$^{18}$Ne~\cite{bib:wie88}, $^{22}$Ne-$^{22}$Mg~\cite{bib:rui03} and $^{46}$Ti-$^{46}$Cr~\cite{bib:hor02,bib:he07} mirror nuclear pairs.
However, a survey of the literature finds inconsistency in the definition of critical parameters, leading to errors in the calculations. 
In the present work a complete and consistent TELD formalism is presented and made available for wider use. 

\section{The Wave Function}
Here we reproduce and expand upon the original work of Thomas~\cite{bib:tho52}, using, for consistency, exactly the same terminology. The channel radius
of two interacting bodies is defined as $a_c=1.44\times(A_{1c}^{1/3}+A_{2c}^{1/3})$ fm, with $A_{1c}$ and $A_{2c}$ being the mass numbers of the bodies
of the pair; the reduced mass is $M_c= A_{1c} A_{2c}/(A_{1c}+A_{2c})$; the energy of relative motion is $\epsilon_c$, which may be positive or negative.
The subscript $c$ is used to describe all of the features of the channel, unless it is necessary to distinguish the positive-energy ($\epsilon_{c+}>0$)
from the negative-energy ($\epsilon_{c-}<0$) channels in which case the symbols $c+$ and $c-$ are used, respectively.

For external wave functions, a radial factor (Equ. 1 of ref~\cite{bib:tho52}) may be written that satisfies the wave equation
\begin{equation}
\overline{F}_c^{\prime\prime} + (2M_c/\hbar^2)(\epsilon_c - \mho_c)\overline{F}_c = 0,
\label{eq:1}
\end{equation}
where a prime signifies differentiation with respect to $r$ (in the following descriptions, all the derivatives are with respect to $r$ unless stated
otherwise). The interaction potential may be written
\begin{equation}
\mho_c = Z_{1c}Z_{2c}e^2r_c^{-1} + (\hbar^2/2M_c)\ell(\ell+1)r_c^{-2},
\label{eq:2}
\end{equation}
where the nuclear potential term disappears in the external region. In the notation of Yost, Wheeler, and Breit~\cite{bib:yos36}, the positive-energy
solution, which is regular at the origin, is designated by ${F(kr)}$ and has the asymptotic form for large $r$,
\begin{equation}
F_{c+} \thicksim \sin(x-\frac{1}{2}\ell \pi -\eta \mathrm{ln}2x+ \sigma).
\label{eq:3}
\end{equation}
Likewise, there is a solution which is linearly independent of $F$ and irregular at the origin which is conveniently taken with the asymptotic form
for large $r$,
\begin{equation}
G_{c+} \thicksim \cos(x-\frac{1}{2}\ell \pi -\eta \mathrm{ln}2x+ \sigma).
\label{eq:4}
\end{equation}
The quantities entering Equ.~\ref{eq:3} and~\ref{eq:4} are
\begin{eqnarray}
x_{c\pm} & = & kr \nonumber \\
k_{c\pm} & = & p/\hbar =(2M_c|\epsilon|/\hbar^2)^{1/2} \nonumber,
\end{eqnarray}
with Sommerfeld parameter
\begin{eqnarray}
\eta_{c\pm}=M_cZ_{1c}Z_{2c}e^2/\hbar^2k =Z_{1c}Z_{2c}e^2/\hbar v \nonumber,
\end{eqnarray}
and
\begin{eqnarray}
\sigma_{c+}=\mathrm{arg}\Gamma(1+\ell+i\eta) \nonumber.
\end{eqnarray}
It is worth noting that $x$ is replaced with $\rho$ in some formulations.

The general solution of this equation, $\overline{F}(r)$, is a linear combination of $F$ and $G$. The Wronskian relation for these two particular
solutions, which directly follows from Equ.~\ref{eq:1}-~\ref{eq:4} is
\begin{equation}
F^\prime G-G^\prime F=k_{c+}.
\label{eq:5}
\end{equation}
Extensive tables~\cite{bib:blo50} and several computer codes~\cite{bib:bar74,bib:bar82,bib:sea02} have been developed for evaluating $F$ and $G$ and
their derivatives when $\eta>0$.

For the $c-$ channels, only the solution to Equ.~\ref{eq:1}, vanishing at large distances from the origin, can occur; it is the
Whittaker function~\cite{bib:whi35,bib:mag48},
\begin{widetext}
\begin{eqnarray}
W_{-\eta,\ell +\frac{1}{2}}(2x_{c-})=
\frac{e^{-x-\eta \mathrm{ln}2x}}{\Gamma (1+\ell+\eta)}\int_0^\infty t^{\ell+\eta}e^{-t}\left (1+\frac{t}{2x} \right) ^{\ell-\eta}dt.
\label{eq:8}
\end{eqnarray}
\end{widetext}
Whittaker function and its derivative may be accurately calculated using the {\tt whittaker\_w}~\cite{bib:nob04} computer code.
However, it is useful to note that if there is no Coulomb interaction in a $c-$ channel, one has from Equ.~\ref{eq:8} for $s$, $p$, $d$, and $f$
orbitals the simpler relations
\begin{eqnarray}
& & W_{0,\frac{1}{2}}(2x)=e^{-x} \label{eq:9} \\
& & W_{0,\frac{3}{2}}(2x)=(1+x^{-1})e^{-x} \label{eq:10} \\
& & W_{0,\frac{5}{2}}(2x)=(1+3x^{-1}+3x^{-2})e^{-x} \label{eq:11} \\
& & W_{0,\frac{7}{2}}(2x)=(1+6x^{-1}+15x^{-2}+15x^{-3})e^{-x} \label{eq:12}
\end{eqnarray}
which can be used for checking the results from a more complicated code.

In discussing conditions at the nuclear surface, one needs to evaluate the real and imaginary parts of the logarithmic derivatives,
$g_c=E^\prime/E$, and these are \cite{bib:tho52},
\begin{eqnarray}
& & g_{c+}^{Re}=(FF^\prime+GG^\prime)(F^2+G^2)^{-1} \label{eq:13} \\ 
& & g_{c+}^{Im}=k(F^2+G^2)^{-1} \label{eq:14} \\
& & g_{c-}^{Re}=W^\prime W^{-1} \label{eq:15} \\
& & g_{c-}^{Im}=0 \label{eq:16} 
\end{eqnarray}
where $g_c=g^{Re}+\mathrm{i}g^{Im}$ and $r_c=a_c$. Although the simple WKB approximation \cite{bib:tho52,bib:mar57} can perform well in calculating the
logarithmic derivatives of the Coulomb and Whittaker functions in specified regions, modern computer codes perform essentially exact calculations and are preferred.
For example, the difference between the WKB approximation and the exact evaluation of $g_{c-}$, performed using the code {\tt whittaker\_w}~\cite{bib:nob04},
in the region 0.1$<x<$5.0, 0.1$<\eta<$10.0 (roughly corresponding to $Z_{1c}Z_{2c}\leq$20 and bound nucleon
energy 0.1$<|E_b|<$10.0 MeV), can be as large as 3\% ($\ell$=1),
1\% ($\ell$=2) and 0.6\% ($\ell$=3), respectively, and becomes
smaller as $\eta$ increases. In the case that $\ell$=0, the differences
can be as much as 10\% when $x<$0.1 and $\eta<$0.9,
or $x<$0.2 and $\eta<$0.5, or $x<$0.5 and $\eta<$0.3. 
In addition, it should be noted that a frequently used subroutine, {\tt COULFG}~\cite{bib:bar82}, was unable to reproduce the results
($F$, $G$ and their derivatives) of the subroutine {\tt RCWFN} near values of $\eta=1.70x+5.13$ unless the
{\tt ACCUR} variable was set to be less than $10^{-17}$ (of course the smaller the safer, {\em e.g.}, $10^{-30}$
if possible). 

\section{Calculation of level displacements}

Under the assumption of charge symmetry of nuclear forces, the $nn$ and $pp$ nuclear interactions are identical. The difference in the excitation energy
of levels in mirror nuclei is therefore due to differences in the Coulomb, electromagnetic spin-orbit, and mass energies. As suggested
by~\cite{bib:tho52}, one can evaluate this by considering $H_p-H_n \equiv V$. Furthermore, in the one-level approximation, irrespective of the
boundary conditions, in the internal region the proton and neutron wavefunctions are the same to within a multiplicative constant.
Under these assumptions, one obtains (Equ. 30a in ref.~\cite{bib:tho52})
\begin{equation}
E_{n}-E_{p}=-\langle V \rangle_\tau + \Delta_{\lambda n} - \Delta_{\lambda p},
\label{eq:18}
\end{equation}
where there is a boundary condition that satisfies $\overline{b}_{nc}=\overline{b}_{pc}$ (defined in Equ. 24c of~\cite{bib:tho52}).
Here, $\langle V \rangle_\tau$ is the mean value of $V$ in the internal region, and $E_n$ and $E_p$ are the Eigenvalues satisfying
$H_{n(p)}\Psi_{n(p)}=E_{n(p)}\Psi_{n(p)}$ of the nucleus with the odd neutron or proton, respectively.
The difference of the level displacements of the neutron and proton states (see Equ. 30b of~\cite{bib:hor02}),
\begin{equation}
\Delta_{\lambda}=\Delta_{\lambda n} - \Delta_{\lambda p} = -\sum_{c\pm}\gamma_c^2(g_{nc}^{Re}-g_{pc}^{Re}),
\label{eq:19}
\end{equation}
is referred to as the boundary condition level displacement.
The energy difference of corresponding levels of mirror nuclei, following from Equ.~\ref{eq:18} and ~\ref{eq:19}, can be written as
\begin{equation}
(E_{n}^{\ast}-E_{n}^{g.s.})-(E_{p}^{\ast}-E_{p}^{g.s.})=\Delta_{\lambda}^{\ast}-\Delta_{\lambda}^{g.s.},
\label{eq:20}
\end{equation}
where it is assumed that the quantity $\langle V \rangle_\tau$ is the same in the excited state as in the ground state. The scripts $\ast$ and {\em g.s.}
denote the corresponding quantities are being evaluated with respect to the excited state and the ground state, respectively. Assuming that the level
displacements of the two ground states are the same simplifies this relation further,

\begin{equation}
E^\ast(n)-E^\ast(p)=\Delta_{\lambda}^{\ast},
\label{eq:21}
\end{equation}
where $E^\ast(n)$ and $E^\ast(p)$ are the corresponding excitation energies in the mirror nuclei.
Therefore, the observed energy difference between mirror nuclear states is due to different level displacements,
$\Delta_{\lambda n}$ and $\Delta_{\lambda p}$, in the two nuclei.

\begin{table*}
\caption{\label{table1} Various definitions of $\Gamma_{\lambda c}$,$\gamma_{\lambda c}^2$, $\theta_{\lambda c}^2$ and
$\theta_{sp}^2$~\cite{bib:lan60,bib:fre60} of a level $\lambda$ in the literature.}
\begin{tabular}{|c|c|c|c|c|}
  \hline
   & $\Gamma_{\lambda c}$ & $\gamma_{\lambda c}^2$ & $\theta_{\lambda c}^2$ & $\theta_{sp}^2$ \\
  \hline
  Thomas~\cite{bib:tho52}          & 2$P_c(\gamma_{\lambda c}^2/a_c)$ & $\frac{\hbar^2}{2M_c a_c}\theta_{\lambda c}^2$    & $a_c\times\frac{u^2(a_c)}{\int_{0}^{a_c}u^2(r)dr}$                    &  \\
  Lane and Thomas~\cite{bib:lan58,bib:lan60} & 2$P_c\gamma_{\lambda c}^2$       & $\frac{\hbar^2}{M_c a_c^2}\theta_{\lambda c}^2$   & C$^2$S$\theta_{sp}^2$ & $\frac{a_c}{2}\times\frac{u^2(a_c)}{\int_{0}^{a_c}u^2(r)dr}$ \\
  French~\cite{bib:fre60}          & 2$P_c\gamma_{\lambda c}^2$       & $\frac{3\hbar^2}{2M_c a_c^2}\theta_{\lambda c}^2$ & C$^2$S$\theta_{sp}^2$ & $\frac{a_c}{3}\times\frac{u^2(a_c)}{\int_{0}^{a_c}u^2(r)dr}$ \\
  \hline
\end{tabular}
\end{table*}

\begin{figure}[thb]
\begin{center}
\includegraphics[width=7cm]{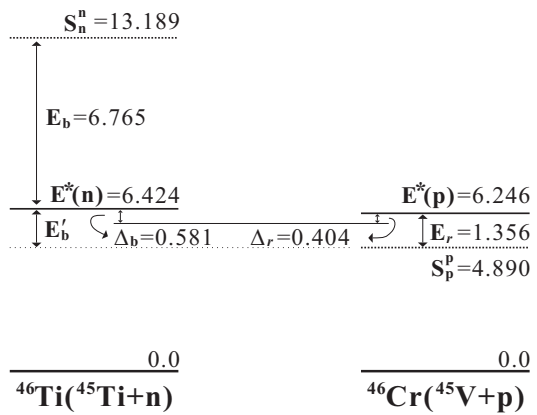}
\end{center}
\caption{\label{prcfig1} Definitions and calculation results of relevant quantities in case of the $^{46}$Ti-$^{46}$Cr mirror pair. The calculated values
(in units of MeV) are with respect to the analogue states at 6.424 MeV (in $^{46}$Ti) and 6.246 MeV (in $^{46}$Cr)~\cite{bib:hjj07}.}
\end{figure}

We find that the definitions used by various authors of partial width $\Gamma_{\lambda c}$, reduced width $\gamma_{\lambda c}^2$, dimensionless reduced
width $\theta_{\lambda c}^2$ and dimensionless single-particle reduced width $\theta_{sp}^2$~\cite{bib:lan60,bib:fre60}, of a level $\lambda$, are not
consistent, as shown in Table~\ref{table1}. Following the previous work, the definitions of French~\cite{bib:fre60} are adopted in the present work,
though it is possible to achieve consistency using the alternative definitions~\cite{bib:tho52,bib:lan58}.
One should pay attention that the definition of $\gamma_{\lambda c}^2$~\cite{bib:fre60} is different from that of~\cite{bib:tho52} by a factor of $3/a_c$.
Thus, for the positive-energy channel,
\begin{equation}
-\gamma_{c+}^2g_{c+}^{Re}=-\frac{3\hbar^2}{2M_c a_c^2}\theta_{c+}^2P_c(FF_{x}^\prime+GG_{x}^\prime)
\label{eq:22}
\end{equation}
with the Coulomb penetrability $P_c=x/(F^2+G^2)$; and for the negative-energy channel,
\begin{equation}
-\gamma_{c-}^2g_{c-}^{Re}=-\frac{3\hbar^2}{2M_c a_c^2}\theta_{c-}^2 x \frac{W_{x}^\prime}{W}.
\label{eq:23}
\end{equation}
Where $F_x^\prime$,$G_x^\prime$ and $W_x^\prime$ represent differentiation with respect to $x$. Equ.~\ref{eq:22} \&~\ref{eq:23} were defined as
$\Delta_b$ and $\Delta_r$ in previous literature by assuming $\overline{b}_{nc}=\overline{b}_{pc}=0$ (see Fig.~\ref{prcfig1}).

The reasonable assumption that the reduced widths, $\gamma_{c}^2$, are the same for the mirror levels leads to an
assumption of $\theta_{c}^2=\theta_{p}^2=\theta_{n}^2$. Thus, the excitation-energy displacements of mirror nuclei can be expressed as
\begin{equation}
\Delta_{\lambda}^{\ast}=\frac{3\hbar^2}{2M_c a_c^2}\theta_c^2 \left\{\left[P_c(FF_{x}^\prime +GG_{x}^\prime)\right]_{\mid E=E_r}- \left(x\frac{W_{x}^\prime}{W}\right)_{\mid E=E_b} \right\},
\label{eq:24}
\end{equation}
where $E_r$ [=$E^\ast(p)-S_p^p$] and $E_b$ [=$|E^\ast(n)-S_n^n|$] are the energies relative to the respective nucleon thresholds ($S_p^p$ and $S_n^n$;
the superscripts $p$ and $n$ refer to the odd-proton (or proton-rich) and odd-neutron (or neutron-rich) nuclei, respectively; the subscripts $p$ and $n$
denote the corresponding nucleon separation energies). A pictorial representation of the physical meanings of the parameters described here are illustrated  
in Fig.~\ref{prcfig1} which shows the case of the 6.424~MeV state in $^{46}$Ti and its analogue state in $^{46}$Cr~\cite{bib:hjj07}. It is clear from the 
figure that the level displacement may equally well be written as,
\begin{equation}
E^\ast(n)-E^\ast(p)=[E^\ast(n)-S_p^p]-[E^\ast(p)-S_p^p]=E_b^\prime-E_r,
\label{eq:25}
\end{equation}
where the quantity $E_b^\prime$ is different from that of $E_b$ defined above. It appears that confusion 
over these definitions has, in part, been the source of errors in the past. For example, in several previous 
works~\cite{bib:lan86,bib:wie88,bib:hor02,bib:rui03} the level displacement has been written as 
$E^\ast(n)-E^\ast(p)=E_b-E_r$ in contrast to the correct expression of $E^\ast(n)-E^\ast(p)=E^\prime_b-E_r$ 
which follows from Marion {\em et al.}~\cite{bib:mar57}. However, the pressent work validates that the correct
expression was used in the calculations~\cite{bib:gor06}.

The final ingredient which is needed to allow calculation of the level shift is the dimensionless reduced width, $\theta_c^2$. 
This can be calculated according to the relation $\theta_c^2$=C$^2$S$\theta_{sp}^2$~\cite{bib:lan60,bib:fre60}, 
where the dimensionless single-particle reduced width $\theta_{sp}^2$ has already been
determined~\cite{bib:ili97,bib:bar98}, and the factor C$^2$S is calculated via a shell-model code such as {\tt OXBASH}~\cite{bib:bro92}. 

\section{Online resources}
\begin{figure}[thb]
\begin{center}
\includegraphics[width=7cm]{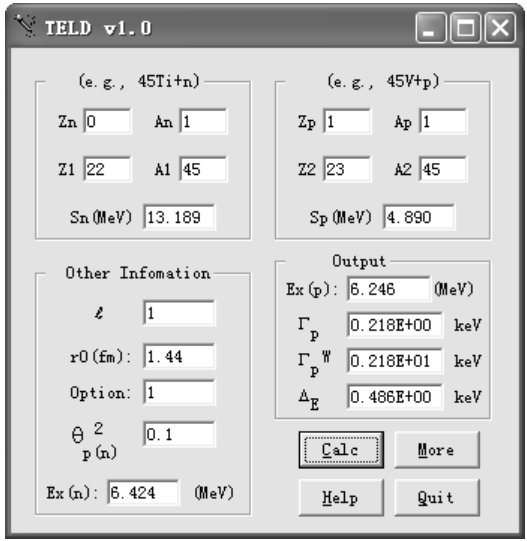}
\end{center}
\caption{\label{VBfig} Screenshot from the {\tt TELD} program for the case of the 6.424~MeV state in $^{46}$Ti and its analogue state 
in $^{46}$Cr~\cite{bib:hjj07}.} 
\end{figure}

The algorithms developed here have been made available online~\cite{bib:heweb}.
The Visual Basic tool requires the {\tt TELD\_VB} and {\tt TELD} executables be in the same directory. Opening the {\tt TELD} 
application provides a window in which one defines the parameters of the mirror states to be considered 
($A$, $Z$, $S_n$, $S_p$ and $E_x(n)$). One also could change values for the channel radius, the orbital angular momentum of the
single particle and spectroscopic factor for the state. 
Activation of the ``calc" button then performs the TELD formalism and returns the resulting excitation of the 
proton-rich analogue state and the proton decay width for this state. This last parameter is calculated using the relation
\begin{equation}
\Gamma_p=\frac{3\hbar^2P_{c}\theta_p^2}{M_c a_c^2}.
\label{eq:29}
\end{equation}
In addition, a Fortran source code, which allows users to make changes whenever required, is also encloded in the TELD.rar package.
Figure~\ref{VBfig} shows a screenshot from the {\tt TELD} program with all relevant parameters for the case of the 
6.424~MeV state in $^{46}$Ti and its analogue state in $^{46}$Cr~\cite{bib:hjj07}. 

\section{Summary}
A complete and consistent Thomas-Ehrman Level Displacement formalism has been presented and made available on the World Wide Web. With this, 
if one has knowledge (or makes a reasonable assumption) of the quantity $\theta_c^2$ and spectroscopic factor $S$, one may estimate the
location of the mirror to a known excited state. Alternatively, experimental measurement of the mirror level displacement
provides a route to determining $\theta_c^2$ (or $S$ factor).

\begin{center}
\textbf{Acknowledgments}
\end{center}

The authors acknowledge support from the Engineering and Physical Sciences Research Council, UK.

\end{document}